\documentstyle[prl,aps,epsfig,multicol]{revtex}
\begin{document}
\draft

\newcommand{\onecolm}{
  \end{multicols}
  \vspace{-1.5\baselineskip}
  \vspace{-\parskip}
  \noindent\rule{0.5\textwidth}{0.1ex}\rule{0.1ex}{2ex}\hfill
}
\newcommand{\twocolm}{
  \hfill\raisebox{-1.9ex}{\rule{0.1ex}{2ex}}\rule{0.5\textwidth}{0.1ex}
  \vspace{-1.5\baselineskip}
  \vspace{-\parskip}
  \begin{multicols}{2}
}

\title{Shot noise for resonant Cooper pair tunneling}
\author{Mahn-Soo Choi$^{(1,2)}$, Francesco Plastina$^{(3)}$,
    and Rosario Fazio$^{(4)}$}
\address{%
        $^{(1)}$Department of Physics and Astronomy, University of Basel,
        Klingelberstrasse 82, 4056 Basel, Switzerland\\
        $^{(2)}$Korea Institute for Advanced Study,
        Cheonryanri-dong 207-43, Seoul 130-012, Korea\\
        $^{(3)}$NEST-INFM \& Dipartimento di Scienze Fisiche ed
        Astronomiche, (DSFA)
        Universit\`a di Palermo, via Archirafi 36, I-90123 Palermo, Italy\\
        $^{(4)}$ NEST-INFM \& Dipartimento di Metodologie Fisiche e
        Chimiche (DMFCI),
        Universit\`a di Catania, viale A. Doria 6, I-95125 Catania,
        Italy
        }
\date{\today}
\maketitle

\begin{abstract}
We study intrinsic noise of current in a superconducting single-electron
transistor, taking into account both coherence effects and 
Coulomb interaction near a Cooper-pair resonance.  Due to this interplay,
the statistics of tunneling events deviates from the Poisson distribution
and, more important, it shows even-odd asymmetry in the transmitted charge.
The zero-frequency noise is suppressed significantly when the quasiparticle
tunneling rates are comparable to the coherent oscillation frequency of
Cooper pairs.
\end{abstract}

%
\newcommand\veps{\varepsilon}
\newcommand\varE{{\mathcal E}}
\newcommand\tr{\mathrm{Tr}}
\newcommand\im{\mathrm{Im}}
\newcommand\lavg{\left\langle}
\newcommand\ravg{\right\rangle}
\newcommand\half{\frac{1}{2}}
\newcommand\ket[1]{\left|#1\right\rangle}
\newcommand\bra[1]{\left\langle#1\right|}

\begin{multicols}{2}

%

Electron tunneling events across a small tunnel junction are correlated
because of the large charging energy.  These correlations lead to variety
of phenomena which fall under the rubric of Coulomb blockade
effects~\cite{singlecharge}. As an important example, the single-electron
transistor (SET) has attracted much interest due to its ultimate
sensitivity to electric charges~\cite{Schoelkopf98}.
If the junctions are superconducting, an additional effect, the coherent
tunneling of Cooper pairs, comes to play and leads to much richer
current-voltage characteristics~\cite{Fulton89a,Averin89}.

Further understanding of the properties of electron transport (related,
e.g., to coherence, electron-electron interaction  and carrier statistics)
comes from study of current fluctuations~\cite{deJong97,Blanter00}.
In single-electron devices, the roles of Coulomb blockade on noise have
been discussed by many authors
\cite{Davies92,Hershfield93,Hanke93-94,Korotkov94a,Sukhorukov00}.
Moreover the importance of coherence, leading to an enhancement of shot
noise in superconducting quantum point contacts, was pointed out in
Ref.~\cite{Averin96a}.
Up to our knowledge, however, the combined effect of coherence and
Coulomb blockade in superconducting double tunnel
junction systems has not been addressed.
Additional interest in studying noise in single-electron devices comes from
their use in quantum measurements~\cite{Makhlin00a,Devoret00} and as
entanglement detectors in solid state systems~\cite{Loss00}.

In this Letter, we discuss the  statistics of tunneling events and the shot
noise in superconducting SET near a resonance for Cooper pair tunneling.
The interplay between coherence and interaction, explored by sweeping
the device through the resonance,  leads to a number of interesting results.
i) At a Cooper pair resonance the statistics of
tunneling events is non-Poissonian and it shows an even-odd asymmetry.
ii) The shot noise suppression depends
strongly on the ratio between the Josephson coupling and the quasi-particle
tunneling rate (the effect is more pronounced close to the resonance).
iii) The frequency-dependent noise has a resonance peak at a frequency
corresponding to the coherent oscillation of Cooper pairs.

The superconducting SET [see Fig.~\ref{noise:fig1} (a)] is  a system of two
small tunnel junctions in series with a small central electrode.
The device operates in the regime in which the
charging energy $E_C=e^2/2C_\Sigma$ ($C_\Sigma$ is the total capacitance of
the island) is much larger than the Josephson coupling energy $E_J$ as well
as the thermal energy $k_BT$. The largest energy scale is the superconducting
gap $\Delta$ (assumed equal in both the electrodes and the island). By
adjusting the bias and gate voltages, $V$ and $V_g$, one can put either the
right or left junction at resonance for Cooper-pair
tunneling~\cite{Averin89,ChoiMS01a}.  We consider the case of resonance
across the left junction.

The effective Hamiltonian\cite{Ingold92} is given by
$H=H_0+H_{qp}+H_T$ with\cite{Blum96}
\begin{equation} \label{noise:H0}
    H_0 = E_C(n+n_0)^2 - eVn_R - E_J\cos2\phi_L \,.
\end{equation}
Here $n\equiv{}n_L+n_R$ is the number of excess
electrons on the central island, $n_L$ ($n_R$) is the number of electrons
that have passed across the left (right) junction \emph{into} the central
electrode, $en_0\equiv{C_RV+C_gV_g}$ is the offset charge on the central
island, and $2\phi_L$  is the superconducting phase difference at the left
junction.  $n_j$ and $\phi_j$ are canonically conjugated
$\left[\phi_j,n_k\right]=i\delta_{jk}$. The terms $H_{pq}$ and $H_T$
describe the quasiparticles on the electrodes and their tunneling across the
junctions, respectively~\cite{Averin89,ChoiMS01a}.
They are given by
\begin{equation} \label{noise:Hqp}
    H_{qp}
    = \sum_{\alpha=L,R,D}\sum_{k\sigma}
    \varepsilon_{k\alpha}\gamma_{k\alpha\sigma}
    ^\dag\gamma_{k\alpha\sigma}^{} \,,
\end{equation}
\begin{equation} \label{noise:HT}
    H_T
    = \sum_{j=L,R}\sum_{kq\sigma}\left[ T_{kq} e^{-i\phi_j}\,
    \gamma_{kj\sigma}^\dag
    \gamma_{qD\sigma}^{} + h.c.\right]
\end{equation}
where $\gamma_{k\alpha}^\dag$ ($\gamma_{k\alpha}$) creates
(annihilates) a quasiparticle with momentum $k$ and energy
$\varepsilon_{k\alpha} = \sqrt{\xi_{k\alpha}^2+\Delta^2}$ in electrode
$\alpha$, $\xi_k$ is the single-particle dispersion, and
$T_{kq}$ is the tunneling amplitude.
Each event of quasiparticle tunneling \emph{into} (\emph{out of}) the
island across the junctions leads to the transition $n\to{n+1}$
($n\to{n-1}$).  The rate is given by
\begin{equation} \label{noise:Gamma}
\Gamma_{L/R}^\pm(n)
= \left[\coth(\beta\varE_{n,\pm}^{L/R})\pm 1\right]
  \frac{\im I_{qp}(\varE_{n,\pm}^{L/R})}{2e} \,,
\end{equation}
where $\varE_{n,\pm}^L=\pm{}E_{n,n\pm1}$,
$\varE_{n,\pm}^R=eV\pm{}E_{n,n\pm1}$, $E_{m,n}=E_C(m-n)(m+n+2n_0)$, and
$I_{qp}$ is related to the quasiparticle tunneling
current~\cite{Werthamer66}.

We will focus on the bias regime $|eV|\simeq 2\Delta+E_C$ ($\gg E_J,
k_BT$) where  two charge states, for example $n=0$ and $n=2$, are nearly
degenerate.  Then due to the strong Coulomb blockade, it suffices to keep
the three charge states, $n=0,1,2$, and two tunneling rates,
$\Gamma_1\equiv\Gamma_R^-(1)$ and $\Gamma_2\equiv\Gamma_R^-(2)$; the other
tunneling rates are negligible.  To simplify the notation, we will  assume
that $\Gamma_1=\Gamma_2\equiv\Gamma$, which is a very good approximation in
the regime we are interested in.  Effectively, one can imagine that across the
left junction only coherent Cooper-pair tunneling occurs, interrupted from
time to time by quasiparticle tunneling across the right junction, see
Fig.~\ref{noise:fig1} (b). 
In the experiment of
Ref.~\cite{Nakamura99}, $1/\Gamma_1=8\mathrm{ns}$ and
$1/\Gamma_2=6\mathrm{ns}$ for
$E_C=2.3E_J=117\mu\mathrm{eV}$ and $\Delta=230\mu\mathrm{eV}$.

We need to keep track of the variable $n_R$ (or alternatively $n_L$) as
well as $n$ ($n_L=n-n_R$). Choosing the basis of $\{\ket{n,n_R}\}$, it can
be shown that only diagonal elements (with respect to $n_R$) of the reduced
density matrix are involved
$
\rho_{mn}(n_R;t) = \bra{m,n_R}\rho(t)\ket{n,n_R} \,.
$
The generalized master equation~\cite{Averin89,Carmichael93} can be written
in the Lindblad form ($\hbar=1$):
\onecolm
\begin{equation} \label{noise:MEq}
\partial_t\rho(n_R)
 =  -i\left[H_0,\rho(n_R)\right]
  + \frac{1}{2}\sum_{n=1,2}\Gamma_n\Bigl[
      2L_n\,\rho(n_R+1)\,L_n^\dag
      - L_n^\dag L_n\,\rho(n_R)
      - \rho(n_R)\,L_n^\dag L_n
    \Bigr]
\label{noise:GME3}
\end{equation}
\noindent\twocolm
where $L_n$ is a Lindblad operator corresponding to the quantum jump
$n\to{n-1}$, $L_n=\ket{n-1} \bra{n}$.  The first term describes a purely
phase-coherent dynamics, while the second is responsible for the
dephasing and relaxation due to the quasiparticle tunneling.

\emph{Counting Statistics\/}.
We first investigate the statistical distribution of the number of
electrons that have tunneled across the right junction.
It has been obtained~\cite{Preparing}, first by defining
the characteristic matrix
\begin{math}
G(\theta,\tau)
= \sum_{n_R}e^{-i\theta n_R}\rho(n_R,\tau+t)
\end{math},
which satisfies a master equation similar to Eq.~(\ref{noise:GME3})
with the initial condition
$G(\theta,0)=\sum_{n_R}\rho(n_R,t\to\infty)$.
The probability $P(N,\tau)$ that $N$ electrons have
tunneled during the period $\tau$ in the stationary state is then given by
\begin{equation} \label{noise:P}
P(N,\tau)
= \int_{-\pi}^\pi\frac{d\theta}{2\pi}\;
  e^{+i\theta N}\tr G(\theta,\tau) \,.
\end{equation}

When the dephasing is strong (either $\Gamma\gg{E_J}$ or
$\veps\equiv{}E_C[(2+n_0)^2-n_0^2]\gg{E_J}$), one can show that
($N<0$)
\begin{mathletters} \label{noise:P2}
\begin{equation} \label{noise:P2a}
P(2N,\tau)  =
\frac{1}{|N|!}\left(\frac{\Gamma_{r}\tau}{2}\right)^{|N|}
  \exp\left(-\frac{\Gamma_{r}\tau}{2}\right)
\end{equation}
\begin{equation} \label{noise:Pbb}
P(2N-1,\tau)
=  0
\end{equation}
\end{mathletters}
where
\begin{math}
\Gamma_{r}
\equiv 2E_J^2\Gamma/(4\veps^2+\Gamma^2)
\end{math} is the relaxation rate
for the charge state population in the strong dephasing limit.
The distribution is Poissonian.  However, there is a strong even-odd
asymmetry.  Physically, the charge is transferred in
pairs (i.e. in units of $2e$) rather than one by one.

In the weak dephasing limit ($\Gamma\ll{E_J}$) at resonance ($\veps=0$),
we find
\begin{mathletters} \label{noise:P1}
\begin{equation} \label{noise:P1a}
P(2N,\tau)
= \exp\left(-\frac{3\Gamma\tau}{4}\right)
  \left(\frac{1}{3}+\frac{4}{\Gamma}\frac{\partial}{\partial\tau}\right)
  F_{|N|}(\tau) \,,
\end{equation}
\begin{equation} \label{noise:P1b}
P(2N-1,\tau)
= \frac{8}{3}\exp\left(-\frac{3\Gamma\tau}{4}\right)\, F_{|N|}(\tau) \,,
\end{equation}
\end{mathletters}
where
\begin{equation}
F_n(\tau)
= \frac{1}{2\pi i}\oint_{|z|=1}\frac{dz}{z^{n+1}}\;
  \frac{1}{\lambda(z)}\sinh\frac{\Gamma\tau\lambda(z)}{4}
\end{equation}
with $\lambda(z)=\sqrt{1+8z}$.
This distribution  shows a much weaker, but still finite, even-odd
asymmetry  than the previous case  [see Eq.~(\ref{noise:P2})].
In the long-time limit
($\Gamma\tau\to\infty$), $P(2N,\tau)=\frac{5}{9}P_G(N,\tau)$ and
$P(2N-1,\tau)=\frac{4}{9}P_G(N,\tau)$ where $P_G(N,\tau)$ is a Gaussian
distribution with $\lavg{N}\ravg=I\tau/2e$ and
\begin{math}
\lavg(\Delta N)^2\ravg = 20\Gamma\tau/27 \,.
\end{math}

In the intermediate case ($\Gamma\sim{E_J}$), an analytic expression for
$P(N,\tau)$ is not available. The numerical results are shown in
Fig.~\ref{noise:fig2}. The distribution function deviates significantly
from a Poissonian distribution function.  Coherent oscillations of the
Cooper pairs manifest themselves in the even-odd asymmetry of the
transmitted charges: $P(N,\tau)$ is suppressed (enhanced) for odd (even)
$N$ compared with the Poissonian distribution.

\emph{Shot Noise\/}. The shot noise spectrum is defined as
\begin{equation} \label{noise:S}
S(\omega) = \int_{-\infty}^\infty{d\tau}\;e^{i\omega\tau}
  \lavg\left\{\delta I(t+\tau),\delta I(t)\right\}\ravg \,,
\end{equation}
where $\delta{I}(t)=I(t)-\langle{I(t)}\rangle$ and $\{A,B\}=AB+BA$.
The total current $I(t)$ through the system is related to the
\emph{tunneling} currents $I_{L/R}=-e\partial_tn_{L/R}$ across the
junctions by
\cite{Korotkov94a}
\begin{equation}
I(t) = \frac{C_R}{C_\Sigma}I_L(t) - \frac{C_L}{C_\Sigma}I_R(t) \;.
\end{equation}
It is convenient to define the spectral densities of tunneling currents
$S_{ij}(\omega)$ ($i,j=L,R$) in an analogous way as in Eq.(\ref{noise:S})
and rewrite the noise power density in the form
\onecolm
\begin{equation}
S(\omega)
= \frac{C_R^2}{C_\Sigma^2}S_{LL}(\omega)
  + \frac{C_L^2}{C_\Sigma^2}S_{RR}(\omega)
  - \frac{C_LC_R}{C_\Sigma^2}\left[S_{LR}(\omega)+S_{RL}(\omega)\right] \,.
\end{equation}
\noindent\twocolm
In the stationary state
$\lavg{I}\ravg=\lavg{I_L}\ravg=-\lavg{I_R}\ravg$,
so that $S(\omega)=S_{LL}(\omega)=S_{RR}(\omega)$ in the
zero-frequency limit.
In the opposite limit ($\omega\to\infty$),
\begin{math}
S(\omega)
= (C_L^2/C_\Sigma^2)S_{RR}(\omega)
= (C_L^2/C_\Sigma^2)\,2e\lavg{I}\ravg
\end{math} \cite{Davies92,Hershfield93,Korotkov94a}.
[In our case, the left junction is (nearly) at resonance for the Cooper pair
tunneling and hence $\lim_{\omega\to\infty}S_{LL}(\omega)=0$; see also the
remarks below Eq.~(\ref{noise:Gamma}).]

In order to calculate the two-time correlators in Eq.~(\ref{noise:S}),
we follow the procedures based on the quantum regression
theorem~\cite{Carmichael93} starting from the master equation
(\ref{noise:MEq}).
An explicit (but lengthy) expression for $S(\omega)$ in terms of
$\Gamma_{1,2}$ and $\veps$ can be given at an arbitrary finite
frequency~\cite{Preparing}. Here we discuss the
zero-frequency shot noise.

At $\omega=0$, the noise power density takes a simple form
\begin{equation} \label{noise:S3}
\frac{S(0)}{2eI}
= 2 - \frac{8(E_J^2+2\Gamma^2)E_J^2}{(3E_J^2 + \Gamma^2 + 4\veps^2)^2} \;.
\end{equation}
In the strong dephasing limit ($\Gamma\gg{E_J}$), the
zero-frequency shot noise in Eq.~(\ref{noise:S3}) is enhanced approximately
by a factor $2$ compared with its classical value, $2eI$.
This can be understood in terms of Josephson quasi-particle (JQP)
cycle~\cite{Fulton89a,Averin89}.  Because of the fast quasiparticle
tunneling across the right junction, each Cooper pair that has tunneled
into the central island breaks up immediately into quasiparticles, and
quickly tunnels out.  The charge is therefore transfered in units of $2e$
(compared with $e$ in classical charge transfer) for each JQP cycle
[see also Eq.~(\ref{noise:P2})].

In the weak ($\Gamma\ll{E_J}$) and moderate ($\Gamma\simeq{E_J}$) dephasing
limits the semiclassical JQP picture breaks down.  In the extreme case
($\Gamma\ll{E_J}$), the quasipaticles do not see the left junction and
consequently the system  can be viewed (approximately) as a single-junction
system.  Still, the noise deviates slightly from the Poisson value since
the channels for tunneling (i.e., $n=1\to0$ and $n=2\to1$ with
corresponding rates $\Gamma_1$ and $\Gamma_2$) are correlated because of
the Cooper pair oscillations and Coulomb blockade.  The effect is related
to the residual even-odd asymmetry of the distribution function in
Eq.~(\ref{noise:P1}).

With moderate dephasing ($\Gamma\simeq{E_J}$), quasiparticle tunneling
events across the right junction are strongly affected by the
\emph{coherent\/} oscillation of Cooper pairs across the left junction.
Indeed, this effect gives rise to the significant deviation from the
Poissonian distribution of the tunneling statistics.  More remarkably, it
also leads to the suppression of the shot noise which is maximum (by factor
$2/5$) at resonance ($\veps=0$) for $\Gamma=\sqrt{2}E_J$, see
Fig.~\ref{noise:fig3}. This is reminiscent of the shot noise suppression in
(non-superconducting) double-junction systems\cite{Hershfield93}, whose
maximal suppression is by factor $1/2$ for the symmetric junctions.  We
emphasize, however, that in the latter case, the coherence was not
essential. In our case, the role of coherence is evident noticing that the
dip in Fano factor [i.e., $S(0)/2eI$] disappears when moving away from the
resonant condition as shown in Fig.~\ref{noise:fig3}.

In Fig.~\ref{noise:fig4} we show the typical behavior of the finite-frequency
noise spectrum in the (a) strong and (b) weak
dephasing limits. It is interesting to notice that (only) in the weak
dephasing limit, there is a resonance peak at $\omega=E_J$.
Near the maximum and for $\Gamma\ll{E_J}$,
the noise behaves like~\cite{Preparing}
\begin{equation} \label{noise:S4}
\frac{S(\omega)}{2eI}
\approx \frac{C_R^2}{2C_\Sigma^2}\,
  \frac{E_J^2}{(\omega-E_J)^2+\Gamma^2/4} \,.
\end{equation}
The peak is an effect of coherent quantum transitions between
the two energy levels tunnel split by $E_J$.

The JQP process discussed in this Letter was used in a recent
experiment~\cite{Nakamura99} to probe the coherent evolution of quantum
states in a Cooper pair box. Weak continuous measurement using quantum
point contact~\cite{Korotkov99a} and strong measurement using
single-electron transistor~\cite{Makhlin00a} have been proposed.  Whereas
both schemes are non-invasive measurements, the setup discussed here
probes the charge states on the island directly and invasively.
In the weak dephasing limit, the resonance peaks in Eq.~(\ref{noise:S4})
and in Ref.~\cite{Korotkov99a} have a similar physical origin, and yet the
latter has a peculiar upper bound.
In the strong dephasing limit, the broad peak around zero frequency
[Fig.~\ref{noise:fig4} (a)] does not fit to the single Lorentzian shape of
Ref.~\cite{Makhlin00a}, which is a manifestation of the random
telegraph noise; nevertheless, $\Gamma_{r}$
and $1/\tau_{\mathrm{mix}}$ in Ref.~\cite{Makhlin00a}
give the same time scale describing the relaxation of population density of
the charge states on the island.

In conclusion, we have investigated the combined effects of coherence and
interaction on the statistics of tunneling events and the shot noise in a
superconducting SET.
It has been shown that the number distribution of tunneled electrons
deviates from the classical Poisson distribution and that zero-frequency
shot noise is suppressed significantly due the coherent oscillation of
Cooper pairs in the presence of Coulomb blockade.

We express special thanks to E. Sukhorukov for many invaluable discussions.
We also acknowledge useful discussions with  Y. Blanter, C. Bruder, G. Falci,
Y. Makhlin, A. Maassen van den Brink, G. Sch\"on, and J. Siewert.

%

%
\begin{figure}\centering
\epsfig{width=.50\linewidth,file=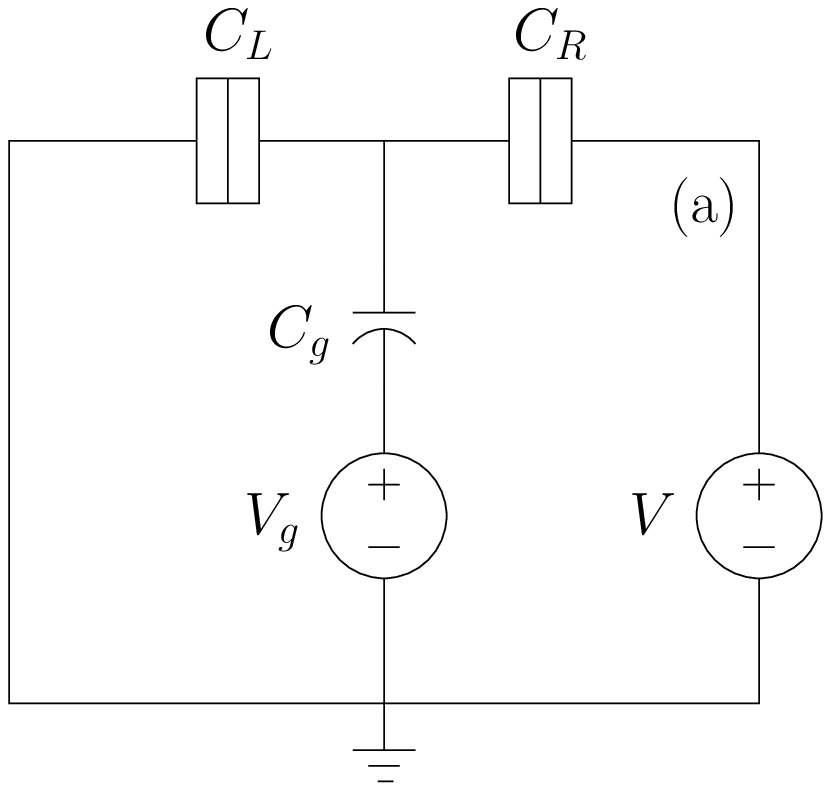}\hfill
\epsfig{width=.45\linewidth,file=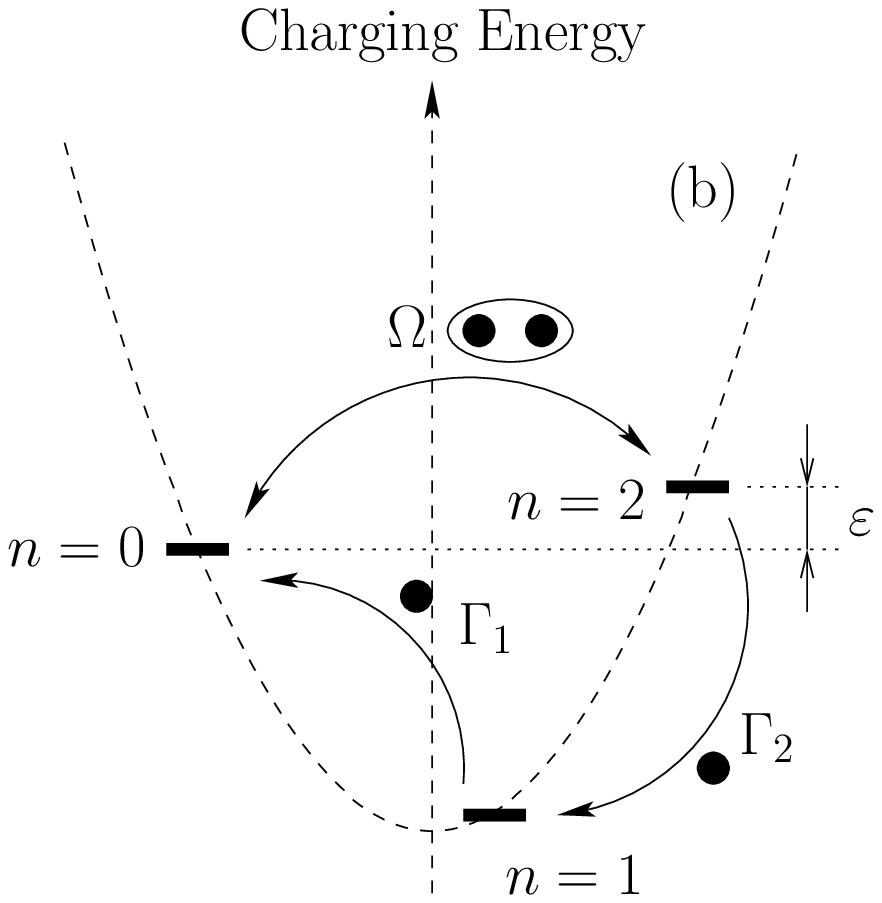}
\caption{Schematic diagrams of (a) the superconducting SET device and (b)
the transition processes between relevant charge states.}
\label{noise:fig1}
\end{figure}

\begin{figure}\centering
\epsfig{width=.59\linewidth,file=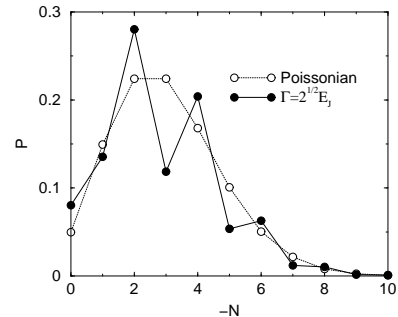}\
\caption{Probability distribution function $P(N,\tau)$ at $\Gamma\tau=4$
for $\Gamma=\sqrt{2}E_J$ (solid line).
For a comparison, the Poissonian distribution is also plotted (dotted
line). Notice that $N<0$ by definition.}
\label{noise:fig2}
\end{figure}

\begin{figure}\centering
\epsfig{width=.59\linewidth,file=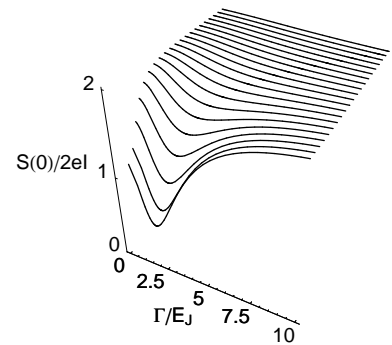}\
\caption{Normalized zero-frequency shot noise for
    $\veps/E_J=0,0.25,\cdots,5$. The dip in the noise is most pronounced at
resonance $(\veps/E_J=0)$.}
\label{noise:fig3}
\end{figure}

\begin{figure}\centering
\epsfig{width=.48\linewidth,file=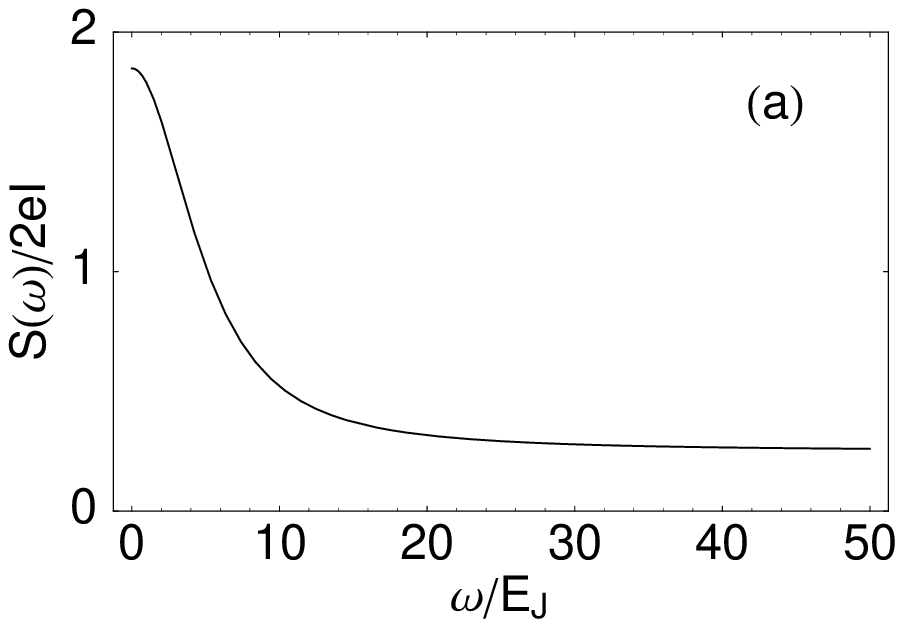}\
\epsfig{width=.48\linewidth,file=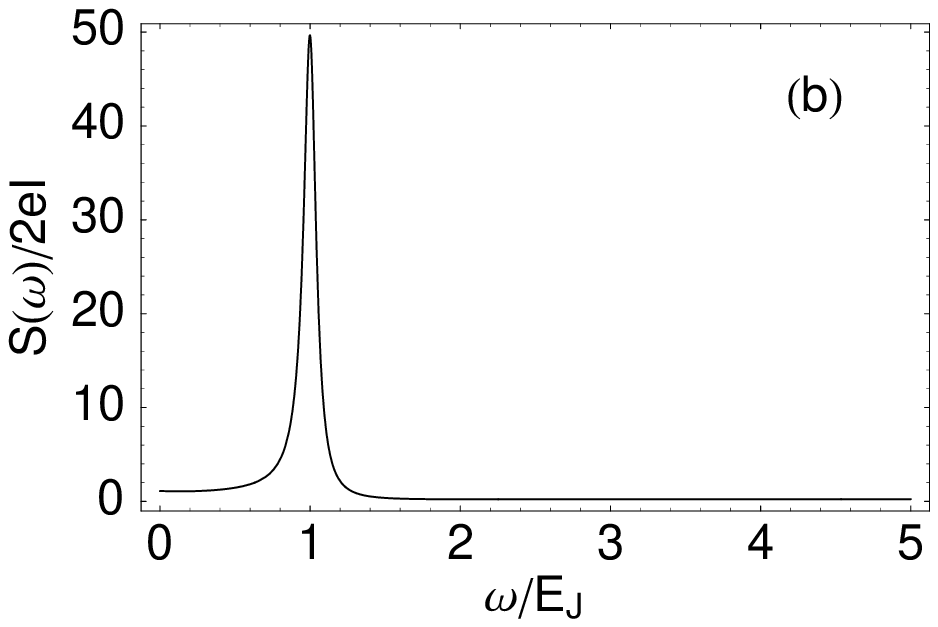}
\caption{Typical behavior of noise power spectrum $S(\omega)$ as a
function of frequency $\omega$ in the (a) strong ($\Gamma_{1,2}\gg{E_J}$)
and (b) weak ($\Gamma_{1,2}\ll{E_J}$) quasi-particle tunneling limits.
For both plots, $C_L=C_R=C_\Sigma/2$ were assumed.}
\label{noise:fig4}
\end{figure}

\end{multicols}

\end{document}